\documentclass[smallextended]{svjour3}       % onecolumn (second format)
\pdfoutput=1
\smartqed  % flush right qed marks, e.g. at end of proof
\usepackage{graphicx}
\usepackage[caption=false,font=footnotesize]{subfig}

\begin{document}

\title{Efficient Floating-Point Givens Rotation Unit%\thanks{Grants or other notes
}


\author{Javier Hormigo         \and
        Sergio~D.~Mu\~noz %etc.
}


\institute{Javier Hormigo \at
              the Department of Computer Architecture,  Universidad de M\'alaga, M\'alaga E-29071 Spain \\
              Tel.: +34-952-132859\\
              Fax: +34-952-132790\\
              \email{fjhormigo@uma.es}           %  \\
           \and
           Sergio~D.~Mu\~noz \at
              the Department of Computer Architecture,  Universidad de M\'alaga, M\'alaga E-29071 Spain \\
              Tel.: +34-952-132859\\
              Fax: +34-952-132790\\
              \email{smunoz@uma.es} 
}

\date{This is a pre-print of an article published in Circuits, Systems, and Signal Processing. The final authenticated
version is available online at: https://doi.org/10.1007/s00034-020-01580-x.}

\maketitle

\begin{abstract}
High-throughput QR decomposition is a key operation in many advanced signal processing and communication applications. For some of these applications, using floating-point computation is becoming almost compulsory. However, there are scarce works in hardware implementations of floating-point QR decomposition for embedded systems. In this paper, we propose a very efficient high-throughput floating-point Givens rotation unit for QR decomposition. Moreover, the initial proposed design for conventional number formats is enhanced by using the new Half-Unit Biased format. The provided error analysis shows the effectiveness of our proposals and the trade-off of different implementation parameters. FPGA implementation results are also presented and a thorough comparison between both approaches. These implementation results also reveal outstanding improvements compared to other previous similar designs in terms of area, latency, and throughput.   
\end{abstract}

\section{Introduction}\label{sec:intro}
Nowadays, most  advanced signal processing  applications, such as  wireless communications, graphics, industrial control and medical imaging, strongly  rely on linear algebra algorithms. Some of these algorithms require performing QR Decomposition (QRD)~\cite{matrixCompu}\cite{Shoaib2013}\cite{Korat2019}. QRD decomposes an input matrix $A^{m\times n}$ into two new matrices, $Q^{m\times m}$ and $R^{m \times n}$, whose product is equal to $A$. Furthermore, it is fulfilled  $R^{m \times n}$ is an upper triangular matrix and $Q^{m\times m}$ is an orthogonal matrix.  QRD or QR factorization is a very compute-intensive operation which requires high-throughput in many applications. Due to that reason, many researchers have investigated the implementation of QRD on hardware for embedded systems. 

There are several methods to compute QRD, but the Givens Rotation Method (and its variations) is probably the most widely used to implement QRD for embedded systems. This is due to its robust numerical properties and its easy parallelization. The Givens Rotation Method is based on a unitary transformation, called Givens rotation, which allows inserting a zero element at a selected location of a matrix. Then, following a predefined schedule, the input matrix is transformed into an upper triangular matrix $R$ by successive Givens rotations, whereas the same rotations over the identity matrix produce an orthogonal matrix $Q$. 

To perform each Givens rotation, first, the rotation angle $\theta$, which allows zeroing an element, has to be computed by using the first non-zero pair of elements of the two target rows. Then, all pairs of elements within said rows have to be rotated by $\theta$. Therefore, at first, the implementation of Givens rotations requires complex logic to compute trigonometric functions. However, most implementations get rid of this complex hardware by using the CORDIC algorithm~\cite{Volder}. Using the same datapath, this algorithm allows computing $\theta$ (vectoring mode) and the vector rotations (rotation mode)  based only on addition and shift operations. 

There are many works on hardware implementation of QRD based on CORDIC architectures, such as in~\cite{5742719}\cite{6936368}\cite{6271699}\cite{Luo2012}\cite{6419866}\cite{TCAS15}\cite{Robles2017}\cite{Korat2019}. 
The majority of these works focus on fixed-point implementation, mostly because Floating-Point (FP) implementations require more resources. However, the increasing complexity of new algorithms produces that the use of FP numbers becomes compulsory for many applications due to either stability or precision requirements. Some of these applications include adaptive
beam-forming~\cite{1193571}\cite{Lightbody200067}\cite{Surapong2011}, Space-Time Adaptive Processing (STAP) for  radar applications~\cite{7015041}\cite{5960667}\cite{Kulikov2020}, adaptive FIR filtering in fetal electrocardiography~\cite{6871780} and 
sub-Nyquist sampling~\cite{7168793}.

In this paper, we study the implementation of a FP Givens rotation unit with high-throughput and low area and energy cost based on the CORDIC algorithm. Our pipelined arithmetic unit focuses only on the computation itself and requires an extremely simple control based on only one signal.  Therefore, our design could be useful for specific applications on embedded systems, and also for being included as computation core in FPGA accelerators designed for high-performance computing applications.

Our design is based on the fixed-point Givens rotator used in~\cite{Luo2012} and~\cite{TCAS15}  which achieves a very high-throughput with a very low cost. That is accomplished by eliminating the z-coordinate datapath, and fully overlapping the angle computation and row-element rotation. 
In this paper, we investigate the adaptation of this fixed-point architecture to support FP numbers with a minimum cost increase. Although there are some proposals for generic FP implementation of CORDIC algorithm, to the best of our knowledge there is no implementation of a specialized FP CORDIC architecture to perform Givens rotations.  Furthermore, in this paper, the proposed  Givens rotator for standard  FP numbers is enhanced by transforming it to support Half-Unit Biased (HUB) FP numbers~\cite{TC15}. Similarly to~\cite{ISCE15QRD}, this new format allows us to reduce simultaneously the area, the delay and the energy consumption of the proposed FP Given rotator.

The main contributions of this paper are:
\begin{itemize}
    \item the proposal of a FP Givens-rotation unit with high throughput and reduced area for IEEE754-like numbers;
    \item the enhancement of the proposed FP Givens rotator by using HUB approach;
    \item an experimental error analysis to optimize the internal word-length and number of CORDIC iterations;
    \item a FPGA implementation analysis and  a comparison of different parameters and approaches;
    \item a comparison with a fixed-point implementation and other previous FP proposals related to QRD.
    
\end{itemize}

The rest of this paper is structured as follows: Section~\ref{sec:review} provides a review of some previous proposals of FP CORDIC architectures and FP Givens rotators. The proposed new architecture to implement the FP Givens rotators is described in Section~\ref{sec:std-arc}. The improvement of this architecture by using HUB format is presented in Section~\ref{sec:hub-arc}. %Section ?? gives some sketches on how utilizes the proposed rotator to implement the QRD. 
The implementation results of the studied architecture along with their error analysis and comparisons are provided in Section~\ref{sec:eval}. Finally, Section~\ref{sec:con} gives the conclusions of this work.

\section{Previous works on FP CORDIC and FP Givens rotation}\label{sec:review}

The CORDIC (COordinate Rotation DIgital Computer) algorithm is an iterative algorithm only based on shifts and additions~\cite{Volder}. It allows rotating an input vector, which is specified by its coordinates $(X, Y)$, through an angle $\theta$, which is usually called $Z$ coordinate. The same CORDIC circuit may operate in two modes, vectoring or rotation mode. The former rotates the input vector until its $Y$ coordinate reaches zero, such as the $Z$ coordinates indicates the angle of the input vector, and the $X$ coordinate, the modulus. The latter rotates the input vector through a specific angle $\theta$. Therefore, a CORDIC unit could be used to compute the angle for a Givens Rotation (vectoring mode), and then performs the rotation through the rest of the row using said angle (rotation mode). 

Since the CORDIC algorithm was first proposed by Volder in~\cite{Volder}, many researchers have proposed different improvements to the algorithm or its implementation, whereas others have utilized them in a wide variety of applications.  One of these improvements is the FP implementation of this algorithm. In~\cite{25655}, a FP CORDIC processor for matrix computation was presented. This word-serial implementation performs each iteration using FP addition and shifting for the $X$ and $Y$ datapath. However, the $Z$ coordinate is represented using fixed-point values with the same bit-width as the significand of the $X$ and $Y$ coordinates. According to the authors, this hybrid processor possesses sufficient accuracy to compute QRD and other matrix computation. A similar approach was used in~\cite{1212843} to implement an SVD processor. Again, an iterative CORDIC architecture is also used, but a two-stage pipeline of the datapath allows performing two independent rotations at the same time.    
 
On the other hand, a FP $Z$ coordinate (angle)  was used by~\cite{378100} and~\cite{Munoz2010} for their word-serial implementation. In the latter, all operations are performed in FP format, which requires a large area occupation, considering it is an iterative architecture. On the contrary, the design in~\cite{378100} only uses a full FP representation for the angle, but the computation of iterations is performed using block FP (i.e., exponents remains constant through the iterations). They state that renormalization of significands between consecutive additions are expensive and not required~\cite{378100}.

In \cite{15343} and \cite{4637696}, a similar solution is utilized for pipeline implementations, because the implementation of FP addition on each step would be unfeasible due to the amount of hardware required. In this case, the inputs and outputs of a fixed-point pipeline CORDIC core are adapted by a floating-to-fixed-point and a fixed-to-FP converter, respectively. In~\cite{15343}, this approach is utilized to implement a CMOS CORDIC processor for  21-bit FP numbers. In~\cite{4637696}, further optimization of the angle datapath allows implementing a high-throughput double-precision FP CORDIC processor on an FPGA. 

All these previous CORDIC approaches have to deal with the implementation of the $Z$ coordinate datapath which, under FP format, is much more complex than the $X$ and $Y$ datapath.  However, the computation of the angle itself is not required to implement a Givens rotation unit, because only knowing the direction of each microrotation in vectoring mode is enough to perform the rotation mode~\cite{Luo2012}. Thus, the use of generic CORDIC implementations instead of a specific Givens rotation unit is very inefficient since an important amount of the circuit is wasted computing unnecessary results. 

There are other CORDIC designs for computation of specific functions in FP format, such as \cite{Nguyen2015}, \cite{Zhu2017} and \cite{Surapong2011}.   In \cite{Nguyen2015} and \cite{Zhu2017}, an architecture to compute only sine and cosine in FP format is proposed. To optimize the design, in the former the angle is introduced in fixed-point format whereas in the latter CORDIC is combined with Taylor expansion and rectangular multipliers. In~\cite{Surapong2011} a FP phase and magnitude digital detector is proposed. However, all these architectures do not allow computing the QRD by themselves.

 Another approach to compute QRD consists in avoiding the use of the CORDIC algorithm and computing the FP Givens rotation using standard FP arithmetic operations as in~\cite{Wang20093}. The work in~\cite{Wang20093} proposes a 2D-systolic array to perform FP computation of QRD in parallel. This approach uses table look-up and Taylor series expansion to implement the FP division and square root required to implement the Givens rotations. As a consequence, the resulting architecture requires a considerable amount of memory and many multipliers which increases the hardware requirements.

In the next section, we propose a FP CORDIC-based pipeline architecture to perform Givens rotations using very reduced hardware. This reduction is achieved because the angle computation and the rotations themselves share the same hardware. Furthermore,  the $Z$ datapath is eliminated and the computation is performed in fixed-point arithmetic since the FP inputs are previously converted into fixed-point and the other way around at the output. Since both operations, angle calculation and rotation, are almost completely overlapped, the pipeline approach allows very high throughput.

\section{FP Givens rotation unit}\label{sec:std-arc}

In this section, we propose a new FP Givens rotation unit based on the pipeline architecture described in~\cite{TCAS15}. All the content of this section is new but Subsection~\ref{sec-fixcordic} which summarizes the core of the architecture in~\cite{TCAS15}. 

\begin{figure} [thb]
\centering
\includegraphics[width=0.50\textwidth ]{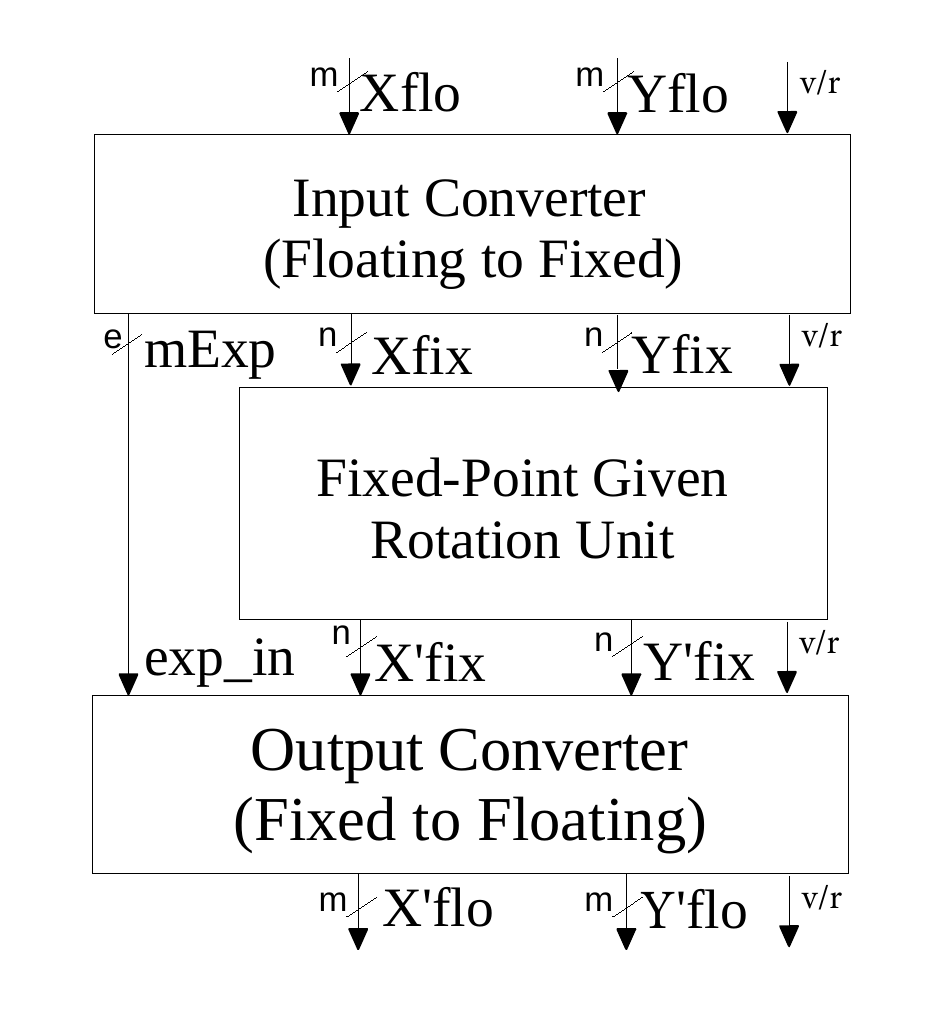}
\caption{General architecture of the proposed FP Givens rotation unit}
\label{ArquiGen_rec}
\end{figure}

As in some previous implementations of FP CORDIC algorithm, here we adapt the fixed-point CORDIC-based architecture to FP numbers by using format converters at its inputs and outputs. These converters do not consider special FP values like NaN, infinity, or subnormals. 
%IF an underflow happens, the value is flush to zero. Similarly, if an overflow occurs while the result is converted to FP, it is saturated to the maximum significand value. 

Fig.~\ref{ArquiGen_rec} shows the general architecture of the proposed FP Givens rotation unit. 
An input converter transforms the FP input coordinates (\textit{Xflo} and \textit{Yflo}) into block FP output coordinates, i.e.  two aligned signed significands (\textit{Xfix} and \textit{Yfix}) sharing the exponent (\textit{mExp}). The aligned significands are two's complement numbers which have one sign bit, one integer bit, and $n-2$ fractional bits. These significands are processed by a fixed-point Givens rotator whereas the exponent is transmitted untouched (except the pipeline delays) through the pipeline to the output converter. The fixed-point Givens rotator performs the desired rotation over the significands producing the fixed-point results (\textit{X'fix} and \textit{Y'fix}). The output converter transforms back the processed block FP results (i.e., the fixed-point output significands along with the common exponent) to independent FP values (\textit{X'flo} and \textit{Y'flo}). The signal \textit{v/r} indicates whether the required operation within the Givens rotator is a vectoring operation for computing the angle or a rotation operation for rotating the corresponding rows.  Following, these three circuits are described in detail.  

\subsection{Input converter for FP to fixed-point conversion}\label{sec-ic}

The input converter of Fig.~\ref{ArquiGen_rec} transforms the two FP values corresponding to the $X$ and $Y$ coordinates into the internal block-FP representation. To do this, the significand of the coordinate with the lowest exponent is aligned by right shifting it, such as this coordinate shares the exponent with the other coordinate. Moreover, both significands are converted from sign-and-magnitude to two's complement representation to facilitate the basic operations required by CORDIC algorithm.% After that, all operations within the CORDIC datapath are performed on the significands in fixed-point arithmetic, whereas the exponent is transmitted unchanged through the pipeline.  

Let us consider that the bit-width of the input significands, $m$, is smaller than the bit-width of the internal significands, $n$. This prevents losing a considerable amount of precision due to the conversion as we will see in Section~\ref{sec:eval}. 

\begin{figure} [tbh]
\centering 
\includegraphics[width=0.75\textwidth ]{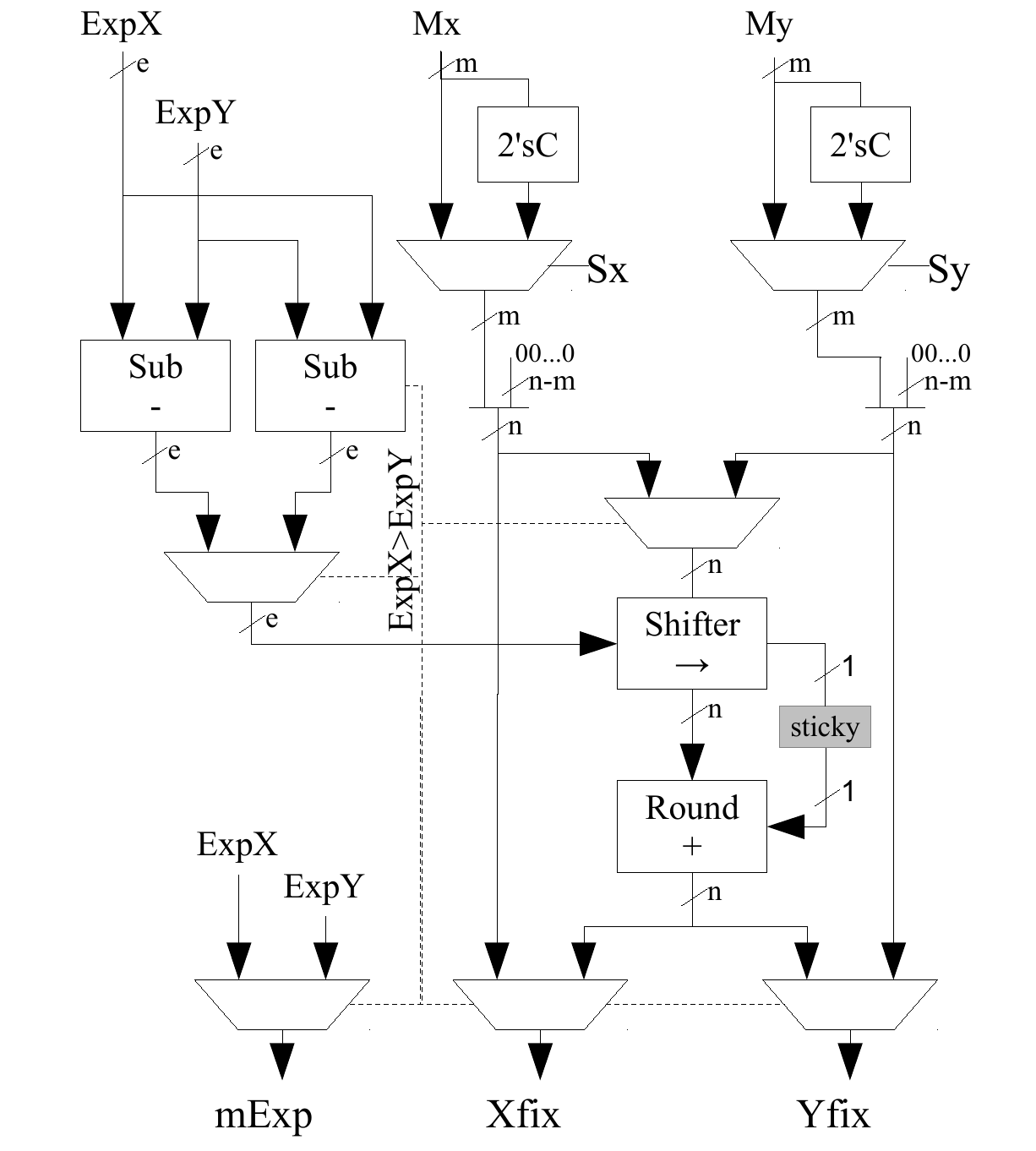}
\caption{ FP to fixed-point converter with rounding.}
\label{FloatToFix_rec}
\end{figure}

Fig.~\ref{FloatToFix_rec} represents the proposed architecture for the  input converter with rounding. The two  FP input values, $X$ and $Y$, are split into sign, exponent, and significand. In Fig.~\ref{FloatToFix_rec}, \textit{Sx}, \textit{ExpX}, and \textit{Mx} represent the sign, the exponent, and the significand  of $X$, respectively, and similarly for the input $Y$. On the other hand, 
the outputs \textit{Xfix} and \textit{Yfix} are the two's complement significands sharing the block exponent \textit{mExp}, obtained from the two inputs values. 

First, the two signed-magnitude input significands, \textit{Mx} and \textit{My}, are converted  into two's complement representation by selecting the two's complement of the value when the corresponding sign bit (\textit{Sx} or \textit{Sy}) is one, and appending the sign bit in the most significant position. Then, the  resulting values  are expanded to fit the required size of $n$ bits by appending $n-m-1$ zeros to the right. 

While the significands are processed,  one input exponent is  subtracted from the other for determining both the absolute difference  between them  and  the greatest one. The two possible subtractions ($ExpX-ExpY$ and $ExpY-ExpX$) are performed in parallel to make the computation faster.   The only positive result of both subtractions is used to select the number of shifting positions for aligning the significand  with the lowest exponent.  The sign of the result of the first subtraction is used  to control the multiplexers which select the output exponent (\textit{mExp}) and the significand for being right shifted. We could implement only one subtraction and use the sign of the result to select the lowest operand, but then a two's complement operation over the difference may be required if its sign is negative. that's mean similar hardware but much lower speed. 

As it is said above, depending on the result of the exponent comparison, the significand  with the lowest exponent is right-shifted by as many bit positions as the absolute difference between the exponents. In the implementation of Fig.~\ref{FloatToFix_rec}, the final value is rounded to nearest tie-to-even~\cite{lang04} based on the discarded bits after shifting to keep the accuracy as high as possible. However, this rounding requires  the computation of the sticky-bit and a possible addition for rounding-up which mean a very significant amount of hardware. Another option is simply discarding the Least Significant Bits (LSBs) of the shifted significand, which simplifies the hardware at the cost of some loss of precision. Both approaches are evaluated in Section~\ref{sec:eval}. 

We should note that Fig.~\ref{FloatToFix_rec} does not represent the circuit to include the leading one of the input significands.  Furthermore, the shifter includes a logic which forces the output to zero if the number of positions  to be shifted is greater than $n$.

\subsection{Fixed-point Givens rotation unit}\label{sec-fixcordic}

To implement the fixed-point Givens rotator, the modified pipeline CORDIC architecture presented in \cite{TCAS15} is used. This pipelined architecture performs both vectoring and rotation mode using the same datapath.  Moreover, it gets rid of the $Z$-coordinate computation by utilizing  directly the direction of each microrotation obtained in the angle computation (vectoring mode) for  rotating the row elements in the next cycles (rotation mode). 

\begin{figure}[tbh]
\centering
\includegraphics[width=0.70\textwidth]{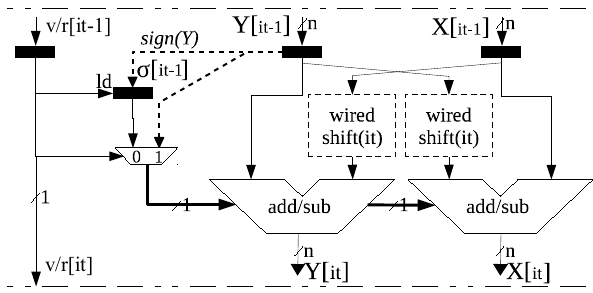}
\caption{Pipelined stage for fixed-point Givens rotation unit}
\label{fig_CORDIC}
\end{figure}

Fig.~\ref{fig_CORDIC} shows a pipeline stage of the  circuit  presented in~\cite{TCAS15}. The right part contains the typical CORDIC  X-Y datapath whereas the left part corresponds to the control logic which substitutes the Z datapath.  In vectoring mode, this circuit selects the microrotation direction indicated by the sign of $Y$ coordinate which is used to control the $add/sub$ circuits in the X-Y datapath. Furthermore, this bit is stored in a register  to be used in the subsequent vector rotations (rotation mode). In the rotation mode, the registers $\sigma$ controls the adders of the X-Y datapath to select the direction of the microrotation. A control signal $v/r$, which is propagated through the pipeline,  is used to select between vectoring or rotation mode. An  active  signal indicates a new angle computation (vectoring mode) on the actual stage. Then, each active $v/r$ is followed by as many inactive cycles (rotation mode) as elements of the row have to be rotated using the computed angle. Therefore, this circuit can perform one element rotation per cycle (either angle computation or row elements rotation) as far as a new pair of row elements are provided at the input each clock cycle. 

\subsection{Output converter for fixed-point to FP conversion}\label{sec-oc}
\begin{figure} [thb]
\centering
\includegraphics[width=0.75\textwidth]{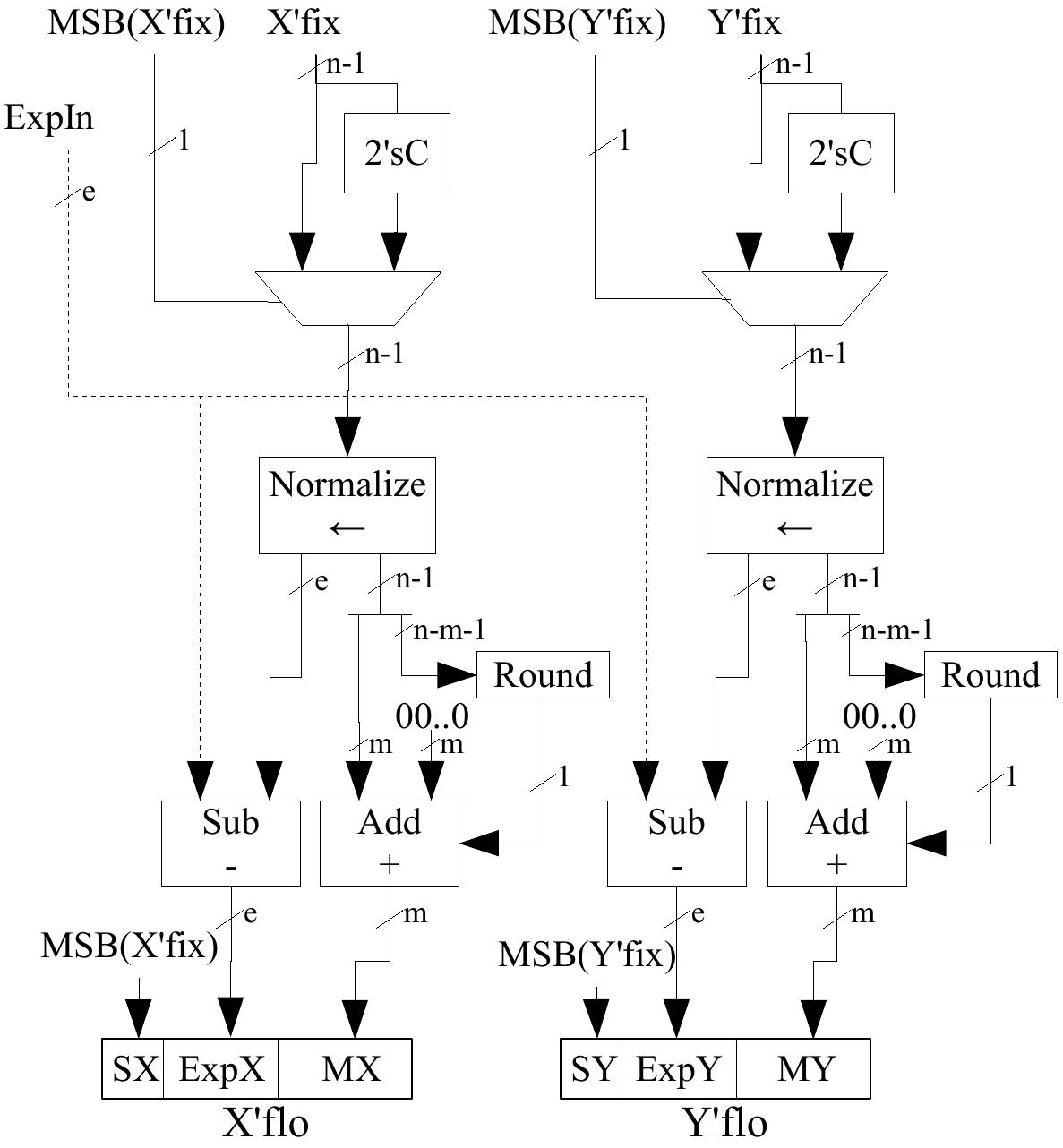}
\caption{Fixed-point to FP converter (under/overflow logic not represented). }
\label{FixToFloat_rec}
\end{figure}

After the CORDIC rotation, the two rotated fixed-point coordinates   have to be converted back to FP representation. This transformation requires the same operations in both fixed-point values: taking the sign bit,  normalizing and rounding the significands, and computing each exponent by actualizing  the common exponent according to said normalization. That could be performed using the architecture shown in Fig.~\ref{FixToFloat_rec}, where the rotated fixed-point coordinates \textit{X'fix} and \textit{Y'fix} are converted into  FP coordinates  \textit{X'flo} and \textit{Y'flo}. 
%which are composed of a sign bit \textit{Sx}, an exponent \textit{ExpX}, and  a significand \textit{Mx}.
 
First, the absolute value of the input coordinates is calculated by the two's complement units and the multiplexer controlled by the Most Significant Bit (MSB) of the corresponding coordinate. This MSB is also the sign of the FP output.  Then, these  unsigned values are normalized  by the  normalization module. This module is composed of a leading one detector and a left-shifter. Using this circuit, the input value is left-shifted until  its MSB equals one. Besides the normalized coordinate, the normalization module also provides the number of shifted positions which is used to compute the corresponding new exponent. This new exponent is calculated by subtracting the number of shifted positions from the common exponent. If the latter operation produces underflow that coordinate is flushed to zero, although for clarity, the logic performing this is not represented in Fig.~\ref{FixToFloat_rec}.   

After normalization, only the $m$ MSBs of the computed value are selected for the output significand. To improve  the accuracy of the FP results, rounding to nearest tie-to-even~\cite{lang04} is utilized. That involves certain logic for computing the sticky-bit and selecting the direction of the rounding based on the discarded bits. Finally, the selected $m$-bit value is incremented by one if a rounding up is required. This rounding may produce an overflow and, in this case, the exponent is incremented accordingly.  

\section{Improvement based on HUB (Half-Unit Biased) approach}\label{sec:hub-arc}

The Half-Unit Biased (HUB) representation is a new family of formats which allows optimizing real number computation by simplifying rounding to nearest and two's complement operation~\cite{TC15}. Basically, HUB formats append an Implicit  Least Significant Bit (ILSB) to the binary number to get the represented value. This ILSB is constant and equals one.  For example, the HUB number $1.0010$ represents the value $1.00101$. When using HUB numbers, rounding to nearest is performed simply by truncation. For example, the nearest 5-bit HUB number to the value $1.101011$ is $1.1010$ (which actually represents $1.10101$), whereas for a conventional representation would be $1.1011$. In this particular example,  the amount of rounding error is the same ($0.000001$) for both cases, but this is not true in the general case.  In fact, it is fulfilled that the addition of the  absolute value of  the rounding error corresponding to both approaches equals the rounding error bound (in this particular example, $0.00001$). That means the better a value is represented under HUB format, the worst it is represented under conventional one. However, the bounds of the rounding errors for conventional and HUB formats are the same~\cite{TC15}. Therefore, although HUB and conventional approaches provide different result values, both representations allow the same accuracy.

On the other hand, another advantage of HUB numbers is the fact that two's complement is performed simply by bit-wise inversion~\cite{TC15}. For example, let consider the signed HUB number $A=01.0110$, then, $-A=10.1001$ (note that the ILSB absorbs the effect of the required increment). This property allows simplifying the implementation of the CORDIC algorithm and the conversion between FP and fixed-point numbers.  

Thanks to those properties, the HUB approach has been very useful to improve both fixed-point or FP designs.  In fixed-point designs, the improvement of accuracy allows reducing the bit-width of numbers and, consequently, also area and delay~\cite{ISCE15QRD}\cite{asil14}. In FP designs, this simplification improves directly the implementation of arithmetic units~\cite{Hormigo2016MeasuringRound-to-Nearest}.   Thus, we propose using this approach to enhance the implementation  of our FP Givens rotation unit. 

 Let us consider that the input and output coordinates in our HUB version of the Givens rotation unit are represented under the HUB FP format. That means the significands has an ILSB whereas exponents remain in conventional format~\cite{TC15}. Similarly, all internal fixed-point significands are also HUB numbers. However, exponent and other auxiliary numbers use conventional representation. Following we show how the circuits described in Section~\ref{sec:std-arc} are turned into a HUB architecture. It would be almost straightforward to adapt this architecture to receive standard FP inputs and deliver HUB FP outputs and the other way around. Combining these three approaches would be very easy to design a QRD unit with standard FP inputs and outputs but working internally with HUB FP numbers.  

\subsection{HUB input converted}\label{sec-hic}
\begin{figure}[thb]
\centering
\includegraphics[width=0.70\textwidth]{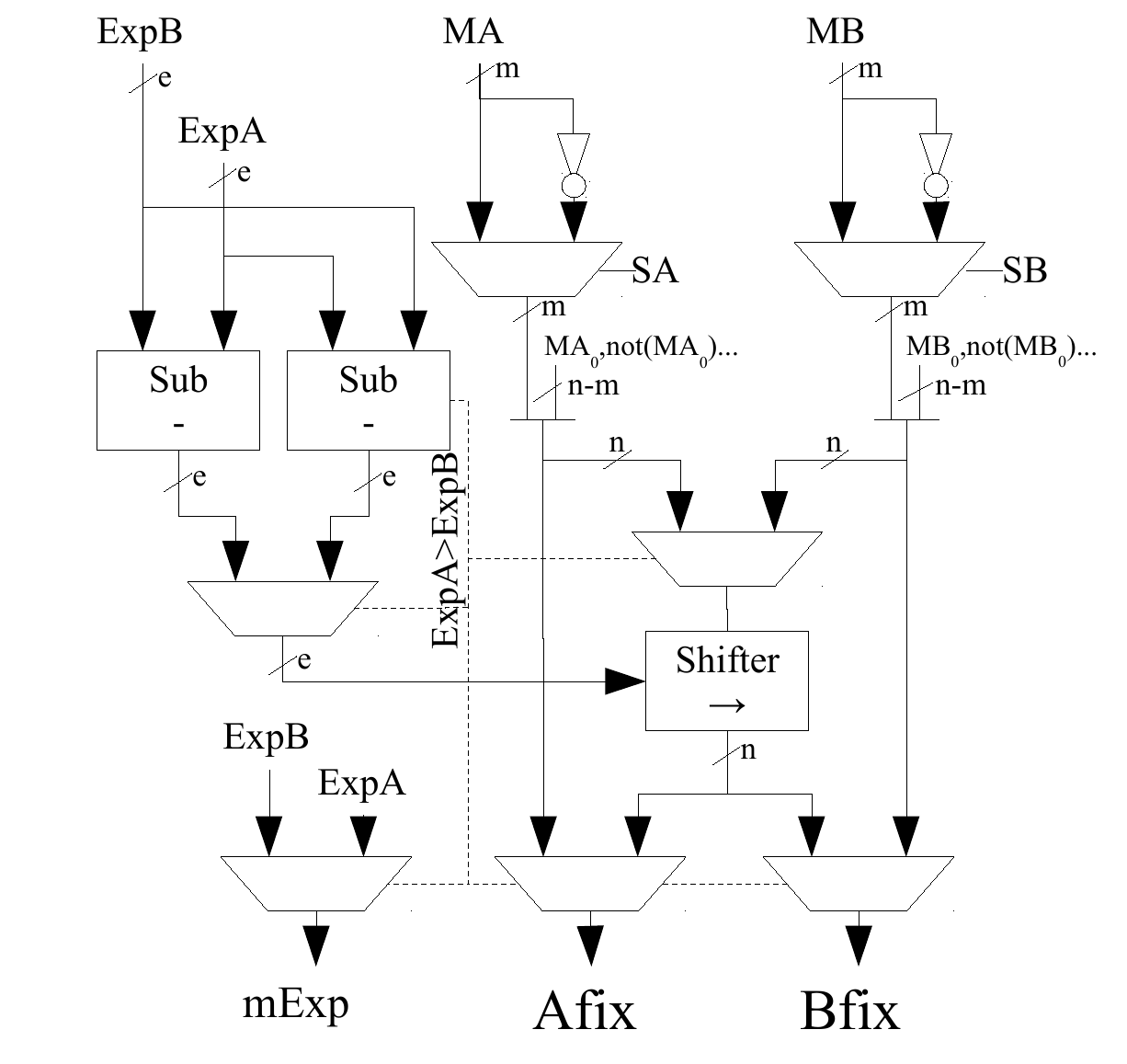}
\caption{FP to fixed-point HUB converter}
\label{FloatToFixHub_rec}
\end{figure}

Here, the input converter described in Section~\ref{sec-ic} is adapted to support HUB numbers, which simplifies it. Fig.~\ref{FloatToFixHub_rec} illustrates the new input converter. First, a simply bit-wise inversion substitutes the two's complement logic in the design of Fig.~\ref{FloatToFix_rec}, since the final addition is not required for HUB numbers. 

Second, the extension of the m-bit significands to reach the $n$ bits requires  appending the ILSB first (which equals one), and then to append the $n-m-1$ zeros. However, the obtained n-bit number is also a HUB number with a new ILSB. That implies an implicit rounding up operation, which may produce some bias in the conversion. To prevent this bias, the extension could be transformed so that the implicit rounding may be either up or down. That could be achieved by extending the number randomly by either '$1000\cdots$' or '$0111\cdots$'~\cite{Hormigo2016MeasuringRound-to-Nearest}. In the architecture of Fig.~\ref{FloatToFixHub_rec} the explicit LSB of the significand has been used as the random variable, such as the significand is extended with this LSB followed by as many bits set to the inverse of the LSB as required to reach the desired bit-width. In Section~\ref{sec:eval}, we compare this approach  with the biased one (zero extension) in terms of precision and hardware cost.  

Third, the main problem of FP HUB formats is the fact that they can not exactly represent integer numbers. In general, that is not a problem in real number computation since integers appear with the same probability as any other real number. Nevertheless, in QR decomposition the identity matrix, which contains 0's and 1's, is introduced as an input if the computation of Q is required. The zeros are not a problem since they are treated as a special number in any case, but when the 1's are managed as a HUB number some error is introduced due to the ILSB. We have studied the introduction of a specific logic to detect the 1's of the identity matrix. If this case is detected, the ILSB of the significand is not appended to the value in the conversions into fixed-point, so that $n-m$ zeros are appended to get an n-bit number. The 1's case is detected by checking that the input exponent is zero (the bits equals '$011\cdots1$' in an IEEE-like representation of the exponent) and the input significand is also zero. In any case, the logic for the latter is included  to detect the zero value  before appending the implicit leading one to the significand. Hence, only the exponent detection has to be added. In Section~\ref{sec:eval}, we  show the pros and cons of using this additional logic.      

Finally, as in the input converter for conventional numbers, after obtaining the two expanded n-bit significands, the one corresponding to the lowest exponent may be right-shifted to align both  input values. In the HUB approach, the obtained shifted value is effectively rounded to the nearest simply by truncation. In contrast to the conventional approach, no additional logic is required for that rounding.  

\subsection{HUB Fixed-point Givens rotation unit}

Although the HUB fixed-point CORDIC-based architecture for QRD was proposed in~\cite{ISCE15QRD}, here, a more detailed description is provided. 

\begin{figure} [thb]
\centering
\includegraphics[width=0.75\textwidth]{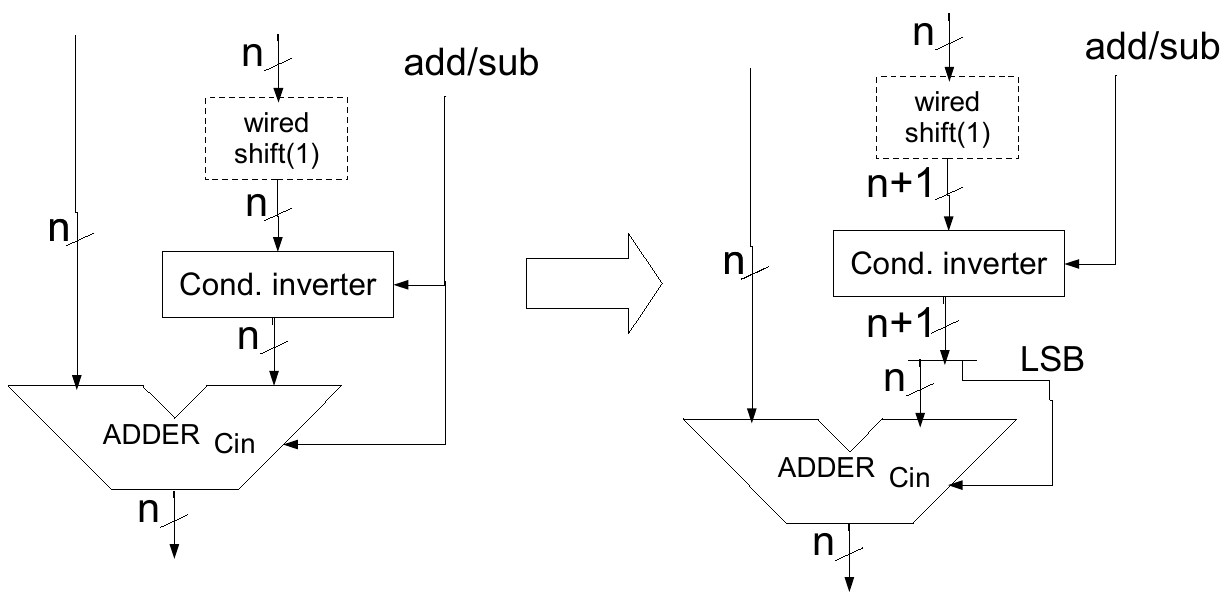}
\caption{Transformation of the CORDIC add/sub circuit for HUB approach}
\label{HUBtrans}
\end{figure}

Since X and Y coordinates are HUB numbers, on each microrotation the ILSBs of both input coordinates have to be considered before the addition/subtraction operation. The introduction of the ILSBs has two significant  implications. First, the two's complement operation utilized for subtraction is simplified since it does not require the addition of the value one. Hence, the input carry  of the adder is available because  it is not set to one for subtraction. Second,  at first, the adder needs one bit more to operate the ILSB appended to the HUB numbers. However, since only the $n$ MSBs of the addition/subtraction are delivered at the output,  the (n+1)th sum bit is not required and only the carry bit has to be actually computed. Taking into account that the (n+1)th MSB of the non-shifted coordinate is always one (since it is the ILSB),  this additional bit of addition could be overcome by connecting the input carry  of the  n-bit adders to the (n+1)th MSB of the shifted coordinate~\cite{ISCE15QRD}. In the first stage,
this bit is one, since it corresponds to the ILSB. Fig.\ref{HUBtrans}  details  the required transformation of each CORDIC stage for the implementation of the addition/subtraction operation.  

Apart from the explicit changes to the circuits, using HUB numbers produces another important effect. Now, the shifted coordinates are not simply truncated but actually rounded before operating them, which produces an increase of the precision. Specifically, in~\cite{ISCE15QRD} is shown that the error in fixed-point QRD computation is halved  by using HUB approach. Thus, HUB implementation could reduce the bit-width by one and keep the same precision as the conventional one.  

\subsection{HUB output converter}

\begin{figure} [thb]
\centering
\includegraphics[width=0.75\textwidth]{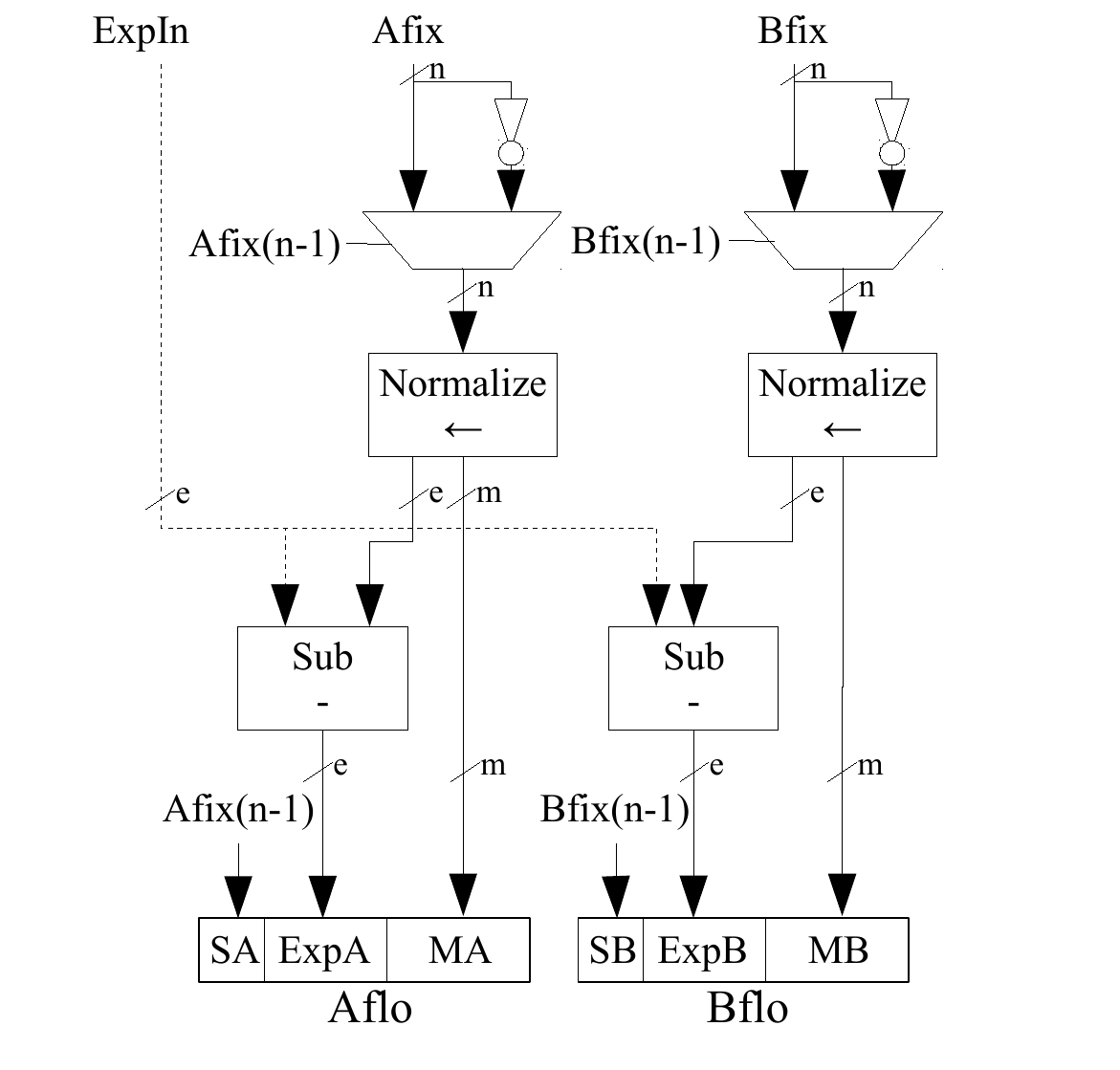}
\caption{Fixed-point to FP HUB converter}
\label{FixToFloatHub_rec}
\end{figure}

Fig.~\ref{FixToFloatHub_rec} shows the circuit of the output converter transformed to support HUB numbers. Similarly to the input converter, a simply bit-wise inversion substitutes the two's complement logic of the previous design to compute the absolute value of the input coordinates. The normalization module is very similar to the conventional one, but the ILSB has to be explicitly appended to the number before left shifting it. However, this solution produces a slightly biased error due to the new ILSB corresponding to the final result. To prevent that bias, in the shifting process the number could be extended randomly with  '$1000\cdots$' or '$0111\cdots$'. As in the HUB input converter, this could be implemented by extending the number with a first bit set to its LSB and the rest using the LSB negated. In the next section, we evaluate both cases, biased and unbiased extension.      

The greatest difference comes after normalization since the HUB implementation discards the $n-m-1$ LSBs instead of utilizing them to compute the round direction. Thus, the HUB output converter gets rid of the round logic which mainly includes the sticky-bit computation logic and the adder (see Fig.~\ref{FixToFloat_rec}). Furthermore, the possibility of having a significand overflow is also eliminated and as a consequence, the increment of the exponent. Those eliminations, along with the two's complement computation, means a very significant reduction in both area and delay.      

\section{Results and comparison}\label{sec:eval}

To analyze the correctness and effectiveness of the proposed architectures, two parametrized Givens rotation units have been implemented in VHDL, one for each approach (conventional and HUB numbers). These designs allow selecting: the bit-width of the floating-point (exponent and significand) and the fixed-point formats, the number of CORDIC microrotations, either rounding or truncation for conventional input converter, and either unbiased or biased extension and identity matrix detection for HUB converters. The Xilinx ISE 14.3 design suite for FPGAs has been used to analyze different aspects of these architectures. 

%We have to note several implementation details. First, the internal CORDIC pipeline appends two integer bit to the fixed-point coordinates to accommodate the value growth due to the scale factor~\cite{lang04}. Second, the scale factor compensation could be performed in the embedded multipliers, but it is not included in the implementation results of the Givens rotators since it is not always necessary.  The output converter has an internal pipeline stage to balance the delay for the stage. 

\subsection{Error analysis}\label{sec:error}
To perform the error analysis, we have used the Monte Carlo method. Our FP Givens rotators are utilized as building blocks to implement a  QRD computation unit for 4x4 matrices following the pipeline architecture proposed in~\cite{TCAS15}. Although the proposed rotator supports any exponent and significand bit-width, only IEEE single-precision FP format (32 bits) has been used in the analysis to simplify it. However, the exponent and the significand bit-width could be adjusted to fit the required dynamic range and the relative precision, respectively.

On each experiment, 10,000 4x4 matrices, with FP values randomly generated in a range bounded by $\pm 2^{\pm r}$ (being $r$ a parameter representing the dynamic range of the input values) are used as inputs. The corresponding Q and R matrices obtained as results of the QRD operation are multiplied ($B=Q^t\times R)$ using double-precision and compared with the original matrix. As error measurement, we use the mean of the Signal-to-Noise Ratio  
(SNR$_{dB}=10\cdot \log_{10}\left( \sum_{i,j} a_{i,j}^2/\sum (a_{i,j}-b_{i,j})^2\right)$,
where $A$ is the input matrix, and $B$ is the matrix obtained) .  As a reference, the same experiments have been carried out using the Matlab function "qr" for single-precision.

\begin{figure} [thb]
\centering
\includegraphics[width=0.75\textwidth]{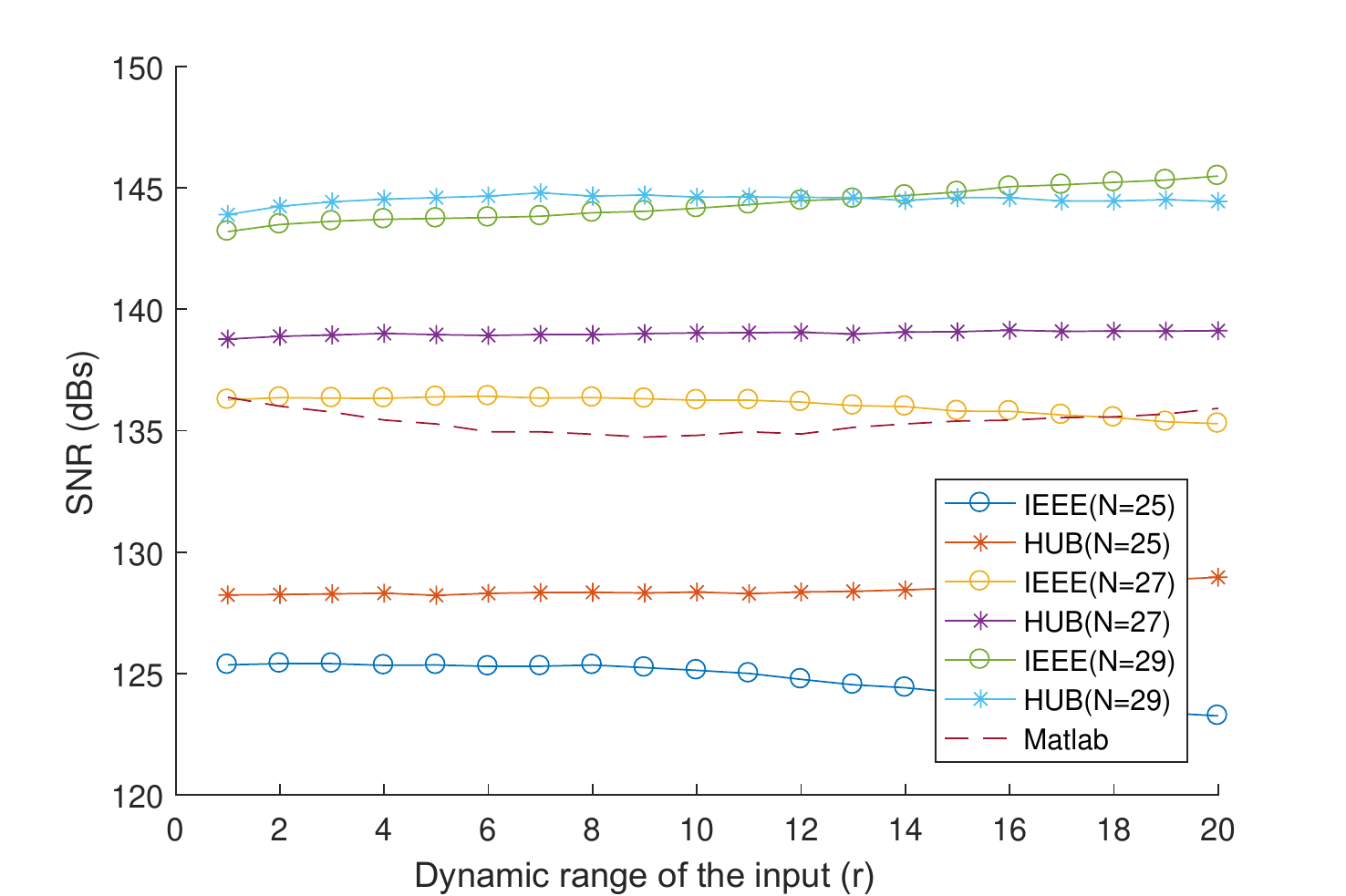}
\caption{Precision of different Givens rotation units when varying $r$ (dynamic range of the input)}
\label{fig-HUBvIEEE}
\end{figure}

As an example, Fig.~\ref{fig-HUBvIEEE} shows the results of the experiments with $r$ ranging between 1 and 20 for both approaches and several fixed-point bit-width ($N={25, 27, 29}$) and $(N-3)$ CORDIC microrotations. The results obtained using Matlab has been included.   First, it is observed that the SNR only change slightly with the dynamic-range parameter  $r$. Hence, we will use the mean of the SNR for all tested values of $r$, since this catches most of the information. Moreover, as expected, the HUB approach performs better than IEEE almost in all cases.

\begin{figure}[tb]
\centering
\subfloat[Conventional approach]{\includegraphics[width=0.75\textwidth]{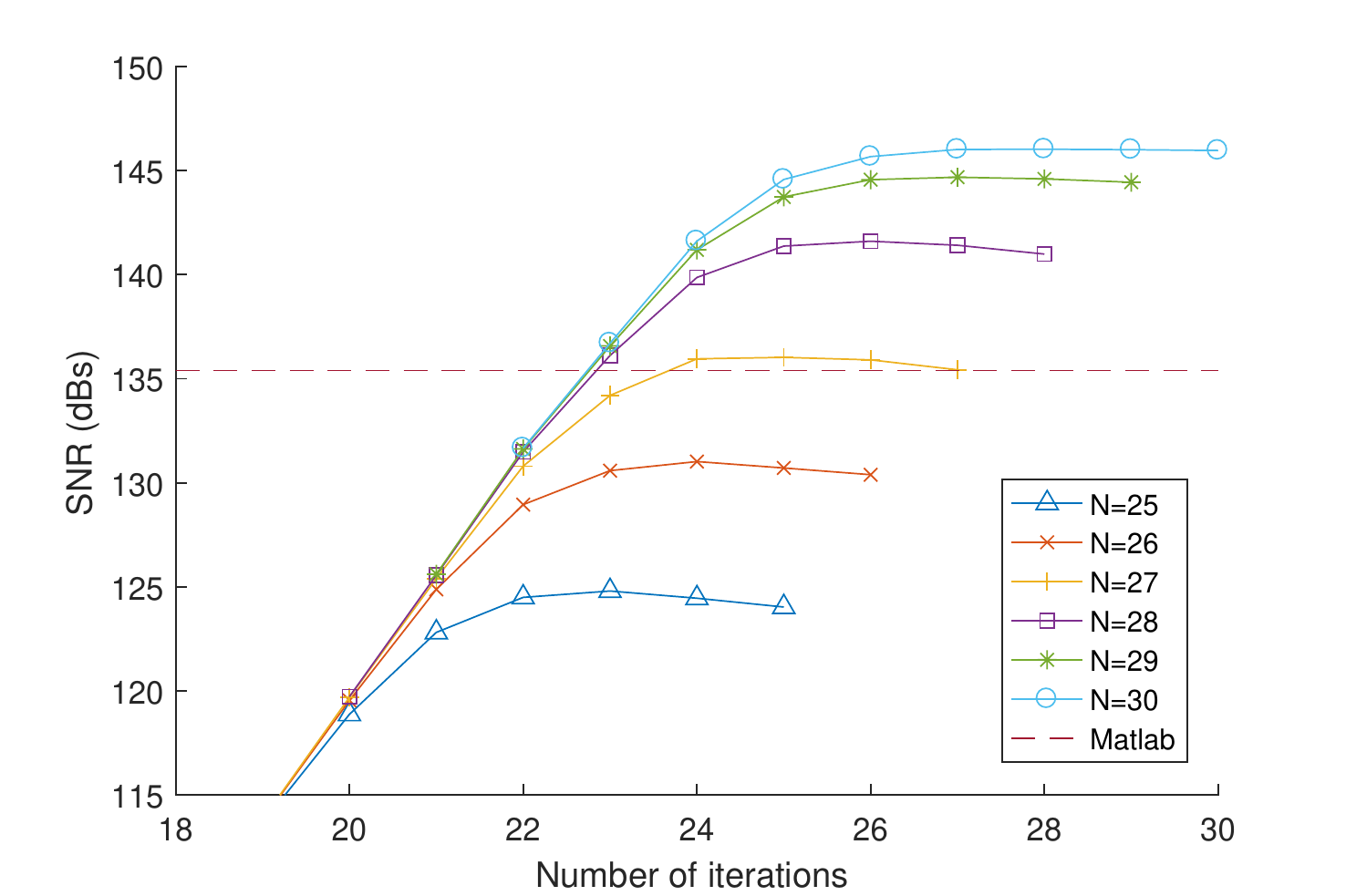}%
\label{fig-iterIEEE}}\\
%\hfil
\subfloat[HUB approach]{\includegraphics[width=0.75\textwidth]{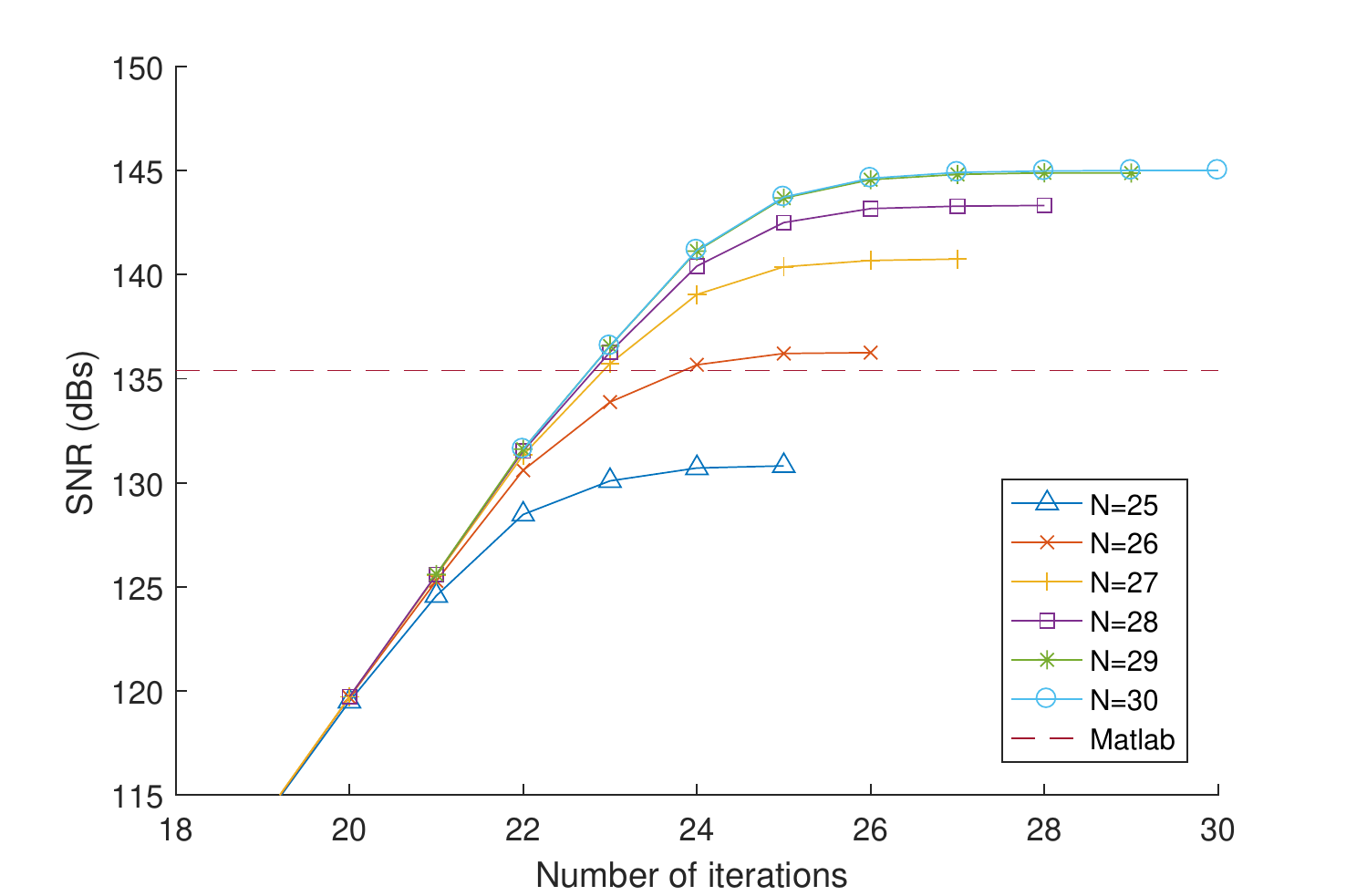}%
\label{fig-iterHUB}}
\caption{Precision achieved when varying the number of CORDIC microrotations for different values of $N$ (internal significand bit-width)}
\label{fig-iter}
\end{figure}

Secondly, to study the ideal numbers of CORDIC microrotations for each case, we have run all the experiments for $N$ ranging from 25 to 30 bits, using different numbers of CORDIC microrotations and $r$ from 1 to 20. In Fig~\ref{fig-iter}, we represent the SNR obtained for each architecture combination. For the conventional approach, using $(N-3)$ microrotations achieves the maximum SNR results and using any more microrotation produces a decrease in the precision. For $N= \{29, 30\}$, $(N-4)$ microrotations achieves almost the same results. Surprisingly, the HUB approach required one microrotation more ($(N-2)$) to reach the top of precision and, in this case, using more microrotations improve very slightly the SNR. It is also clearly observed that the internal fixed-point numbers of the HUB rotators require one bit less than its conventional counterpart to reach the same precision. For the HUB architecture using as less as two guard bits ($N=26$) is enough to reach the same precision as Matlab. Furthermore, $N=29$ and $N=30$ reach the same precision which is the precision of a single-precision FP number. That is why it is not possible to go farther.

\begin{figure} [tb]
\centering
\includegraphics[width=0.75\textwidth]{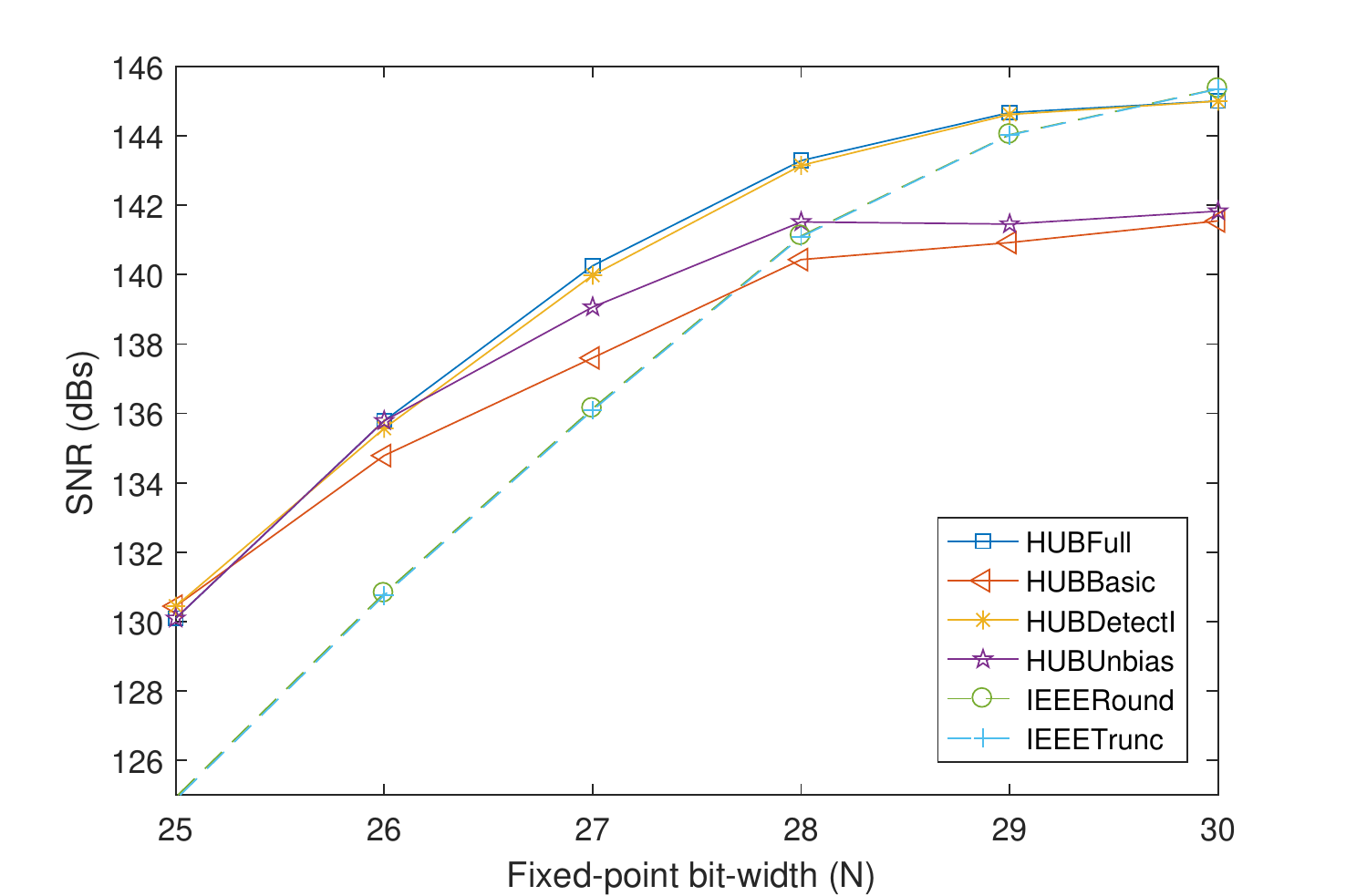}
\caption{Precision of different approaches when varying $N$ (internal significand bit-width)}
\label{fig-mejoras}
\end{figure}

Finally, to analyze the effectiveness of some design proposals, we have run the experiment for different versions of the same architecture, specifically, for IEEE approach, the input converter with truncation (IEEETrunc) and rounding (IEEERound); and for the HUB approach, unbiased version of the converters and detection of the identity matrix ($I$) (HUBFull), only unbiased (HUBunbias) or only $I$ detection (HUBDetectI), and the basic architecture with biased converters and no detection (HUBBasic). Fig.~\ref{fig-mejoras} shows the different results obtained when varying $N$ (the SNR is the mean of the values obtained for $r$ between 1 and 20). For the IEEE versions, it is clear that using rounding in the input converter does not improve the results. On the contrary, $I$ detection enhances the precision of HUB approaches up to 4 dB whereas unbiased conversion only has a significant impact when $I$ detection is not implemented.    
 
\subsection{Implementation results}

Before presenting the implementation results, we have to note some details of the implemented circuits. First, the internal CORDIC pipeline appends two integer bit to the $N$-bit fixed point coordinates to accommodate the growth of the value due to the scale factor~\cite{lang04}. Second,  the scale factor compensation could be performed in the embedded multipliers but it is not included in the implementation results of the Givens rotators since it is not always necessary. Finally, the converters have been pipelined to balance their delay with the CORDIC stages. Specifically, the input converter has two stages whereas the output converter has three. 

Both proposed approaches of the Givens rotation unit (IEEE, and HUB version), have been synthesized using Xilinx ISE 14.3 software for a wide range of configurations targeting a Virtex-6 XV6VLX240T-2 FPGA.   Here, we summarize only the most relevant results obtained using this software tool.  
 
First, Table~\ref{tab:DIvsH}, Table~\ref{tab:AIvsH}, and Table~\ref{tab:PIvsH} allow us to compare the implementation results of both approaches for the most typical FP sizes. For a fair comparison, the internal fixed-point bit-width ($N$) has been selected such as both approaches achieve similar precision. Therefore, according to  Section~\ref{sec:error}, the HUB version uses one bit less than the IEEE one and both have the same number of CORDIC stages (see Fig.~\ref{fig-iter}). Furthermore, the IEEE version uses truncation in the input converters and HUB version uses identity matrix detection and unbiased extension.  Although we show only some concrete values of $N$ (to reach a wider range), relative values are similar for other sizes. The power and energy consumption per operation have been estimated using Xilinx XPS supposing the units work at maximum speed.   Along with the obtained results, the ratio between both approaches (HUB/IEEE) has been included to facilitate the comparison. Clearly, the HUB approach outperforms IEEE one in delay, area, and energy consumption. Using practically the same the number of registers, HUB format reduces the number of LUTs used between 7\% and 18\% and the critical path delay between 24\% to 33\%. IEEE versions require much less power due to the lower frequency, but they consume slightly more energy per operation than the HUB ones (between 3\% and 7\%).

% \begin{table*}[thb]
% \renewcommand{\arraystretch}{1.2}  
% \caption{Implementation results for Givens rotation units in Virtex-6}
%     \label{tab:IvsH}   
% \centering
%     \begin{tabular}{|c||cc|cc|c|cc|c|cc|c|cc|c|cc|c|}
%   \hline
%       & \multicolumn{2}{c|}{$N$} &\multicolumn{3}{c|}{Delay (ns)}&\multicolumn{3}{c|}{Area(LUTs)}&\multicolumn{3}{c|}{Area (Registers)}&\multicolumn{3}{c|}{Power (W)}&\multicolumn{3}{c|}{Energy (pJ)}\\ \hline
%  FP &  IEEE&	HUB&IEEE&	HUB&ratio&IEEE&	HUB&	ratio&IEEE&	HUB&ratio&IEEE&	HUB&	ratio&IEEE&	HUB&ratio\\ \hline
% Half    &14&	13&	2.863&	2.18 &	0.76&	839	&	689	&	0.82&	536	&	513	&	0.96&	0.068&	0.085&	1.24&	195.1 &	184.5 &	0.95\\
% &16&	15&	3.134&	2.315&	0.74&	1030&	825	&	0.80&	680	&	645	&	0.95&	1043&	0.091&	1.26&	225.1 &	209.7 &	0.93\\\hline
% Single  &26&	25&	3.306&	2.337&	0.71&	2365&	2057&	0.87&	1632&	1587&	0.97&	0.131&	0.178&	1.36&	434.0 &	415.8 &	0.96\\
% &28&	27&	3.373&	2.458&	0.73&	2631&	2300&	0.87&	1856&	1845&	0.99&	2654&	0.189&	1.33&	478.9 &	464.1 &	0.97\\
% &30&	29&	3.463&	2.678&	0.77&	2957&	2550&	0.86&	2134&	2060&	0.97&	2985&	0.190&	1.23&	534.4 &	508.1 &	0.95\\\hline
% Double  &55&	54&	4.355&	2.932&	0.67&	8052&	7400&	0.92&	6484&	6461&	1.00&	0.331&	0.481&	1.45&	1440.8&	1409.1&	0.98\\
% &57&	56&	4.65 &	2.865&	0.62&	8508&	7766&	0.91&	6960&	6853&	0.98&	8548&	0.518&	1.57&	1530.4&	1483.4&	0.97\\
% &59&	58&	4.506&	2.999&	0.67&	9012&	8226&	0.91&	7426&	7313&	0.98&	9054&	0.525&	1.46&	1622.7&	1573.0&	0.97\\\hline
%     \end{tabular}
% \end{table*}

\begin{table}[thb]
\caption{Critical path for Givens rotation units in Virtex-6}
    \label{tab:DIvsH}   
\centering
    \begin{tabular}{llllll}
   \hline\noalign{\smallskip}
      & \multicolumn{2}{l}{$N$} &\multicolumn{3}{l}{Delay (ns)}\\ 
 FP &  IEEE&	HUB&IEEE&	HUB&ratio\\ 
 \noalign{\smallskip}\hline\noalign{\smallskip}
    
Half    &14&	13&	2.863&	2.18 &	0.76	\\
		&16&	15&	3.134&	2.315&	0.74	\\
\noalign{\smallskip}\hline\noalign{\smallskip}
Single  &26&	25&	3.306&	2.337&	0.71	\\
		&28&	27&	3.373&	2.458&	0.73	\\
		&30&	29&	3.463&	2.678&	0.77	\\
\noalign{\smallskip}\hline\noalign{\smallskip}
Double  &55&	54&	4.355&	2.932&	0.67	\\
		&57&	56&	4.65 &	2.865&	0.62	\\
		&59&	58&	4.506&	2.999&	0.67	\\
\noalign{\smallskip}\hline
    \end{tabular}
\end{table}

\begin{table}[thb]
\caption{Area results for Givens rotation units in Virtex-6}
    \label{tab:AIvsH}   
\centering
    \begin{tabular}{lllllllll}
   \hline\noalign{\smallskip}
      & \multicolumn{2}{l}{$N$} &\multicolumn{3}{l}{Area(LUTs)}&\multicolumn{3}{l}{Area (Registers)}\\ 
 FP &  IEEE&	HUB&IEEE&	HUB&ratio&IEEE&	HUB&	ratio\\ \noalign{\smallskip}\hline\noalign{\smallskip}
    
Half    &14&	13&		839	&	689	&	0.82&	536	&	513	&	0.96\\
		&16&	15&		1030&	825	&	0.80&	680	&	645	&	0.95\\
\noalign{\smallskip}\hline\noalign{\smallskip}
Single  &26&	25&		2365&	2057&	0.87&	1632&	1587&	0.97\\
		&28&	27&		2631&	2300&	0.87&	1856&	1845&	0.99\\
		&30&	29&		2957&	2550&	0.86&	2134&	2060&	0.97\\
\noalign{\smallskip}\hline\noalign{\smallskip}
Double  &55&	54&		8052&	7400&	0.92&	6484&	6461&	1.00\\
		&57&	56&		8508&	7766&	0.91&	6960&	6853&	0.98\\
		&59&	58&		9012&	8226&	0.91&	7426&	7313&	0.98\\
\noalign{\smallskip}\hline
    \end{tabular}
\end{table}

\begin{table}[thb]
\caption{Power consumption for Givens rotation units in Virtex-6}
    \label{tab:PIvsH}   
\centering
    \begin{tabular}{lllllllll}
   \hline\noalign{\smallskip}
      & \multicolumn{2}{l}{$N$}&\multicolumn{3}{l}{Power (W)}&\multicolumn{3}{l}{Energy (pJ)}\\ \hline
 FP &  IEEE&	HUB&IEEE&	HUB&ratio&IEEE&	HUB&	ratio\\ 
 \noalign{\smallskip}\hline\noalign{\smallskip}
    
Half    &14&	13&	0.068&	0.085&	1.24&	195.1 &	184.5 &	0.95\\
&16&	15&	1043&	0.091&	1.26&	225.1 &	209.7 &	0.93\\
\noalign{\smallskip}\hline\noalign{\smallskip}
Single  &26&	25&	0.131&	0.178&	1.36&	434.0 &	415.8 &	0.96\\
&28&	27&	2654&	0.189&	1.33&	478.9 &	464.1 &	0.97\\
&30&	29&	2985&	0.190&	1.23&	534.4 &	508.1 &	0.95\\
\noalign{\smallskip}\hline\noalign{\smallskip}
Double  &55&	54&	0.331&	0.481&	1.45&	1440.8&	1409.1&	0.98\\
&57&	56&	8548&	0.518&	1.57&	1530.4&	1483.4&	0.97\\
&59&	58&	9054&	0.525&	1.46&	1622.7&	1573.0&	0.97\\
\noalign{\smallskip}\hline
    \end{tabular}
\end{table}

Since there is certain regularity in the implementation results, Table~\ref{tab:vari} condenses the mean area cost of different variations of the architectures presented in Table~\ref{tab:AIvsH}. This table shows only the area increment since delay variations are much lower and less predictable. First and second columns present this relative increment when increasing by one the number of CORDIC microrotations and the fixed-point bit-width ($N$), respectively. Note that increasing $N$ also means increasing the number of microrotations to take full advantage of the higher precision.  
   Third and fourth columns show the area increment  in the HUB version when using the unbiased extension approach and Identity matrix detection, respectively. Taking into account the precision improvement versus the area cost, identity matrix detection is worth implementing when the computation of Q is required. However, the implementation of the unbiased extension seems less likely to be worthwhile.  

\begin{table}[thb]
\caption{Relative area cost when modifying the design parameters}
    \label{tab:vari}   
\centering
    \begin{tabular}{lllllll}
   \hline\noalign{\smallskip}
        &\multicolumn{2}{c}{microrotation}&\multicolumn{2}{c}{$N$}&Unbiased& I Detection\\ 
 FP & IEEE&	HUB&IEEE&	HUB& HUB& HUB\\ 
 \noalign{\smallskip}\hline\noalign{\smallskip}
Half   &	4.4\%&	5.3\%&	10.0\%&	12.8\%	&	0.3\%&	1.0\%	\\
Single &	3.1\%&    2.8\%&	5.3\%&	6.0\%&	2.0\%&0.3\%	\\
Double &   1.4\%&	1.6\%&	3.1\%&	3.1\%&	0.2\%&	0.1\%\\
\noalign{\smallskip}\hline
    \end{tabular}
\end{table}

\subsection{Comparison with fixed-point rotators}

\begin{figure}[thb]
\centering
\subfloat[]{\includegraphics[width=0.75\textwidth]{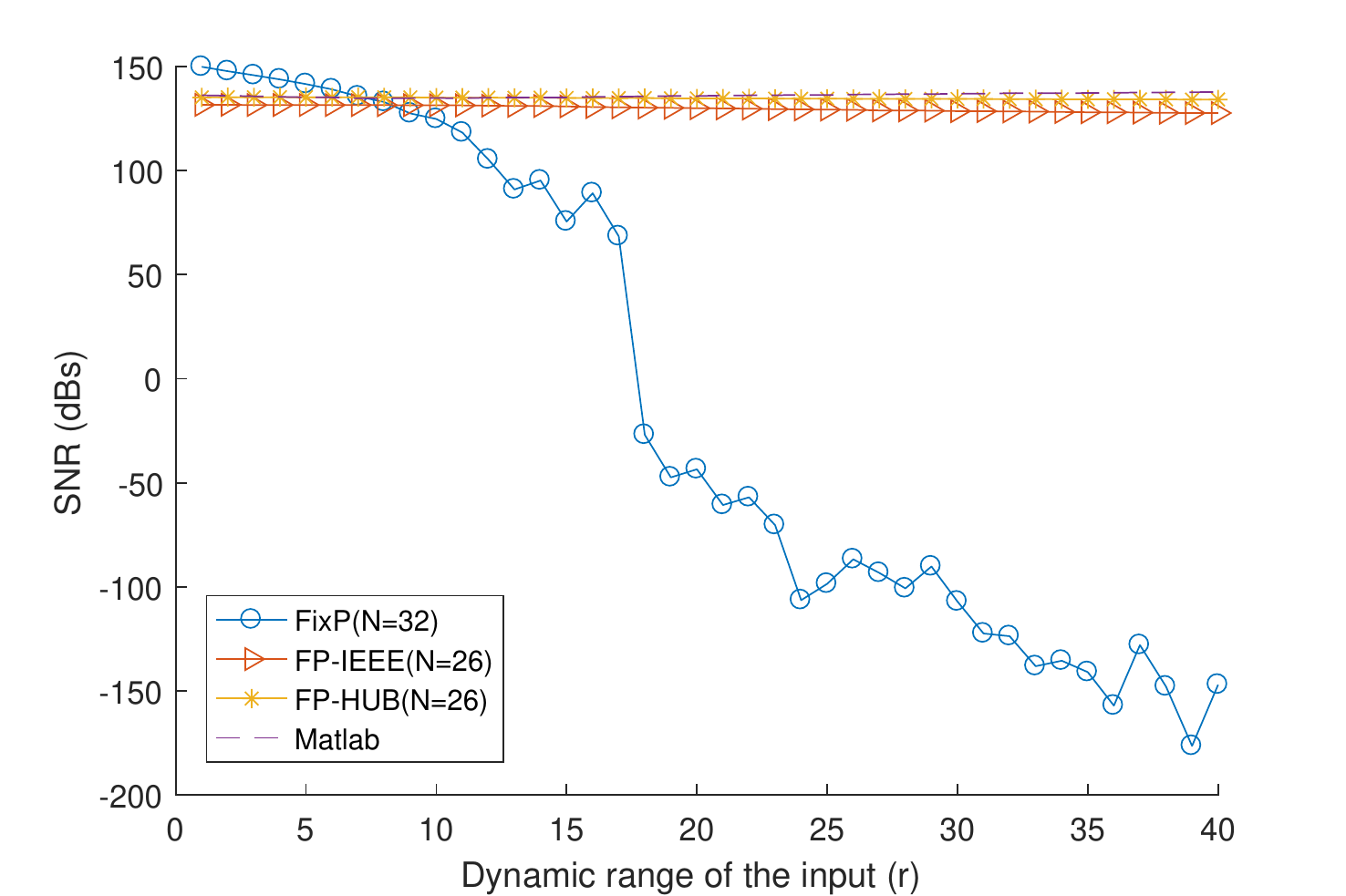}%
\label{fig-fixvsfp40}}\\
%\hfil
\subfloat[]{\includegraphics[width=0.75\textwidth]{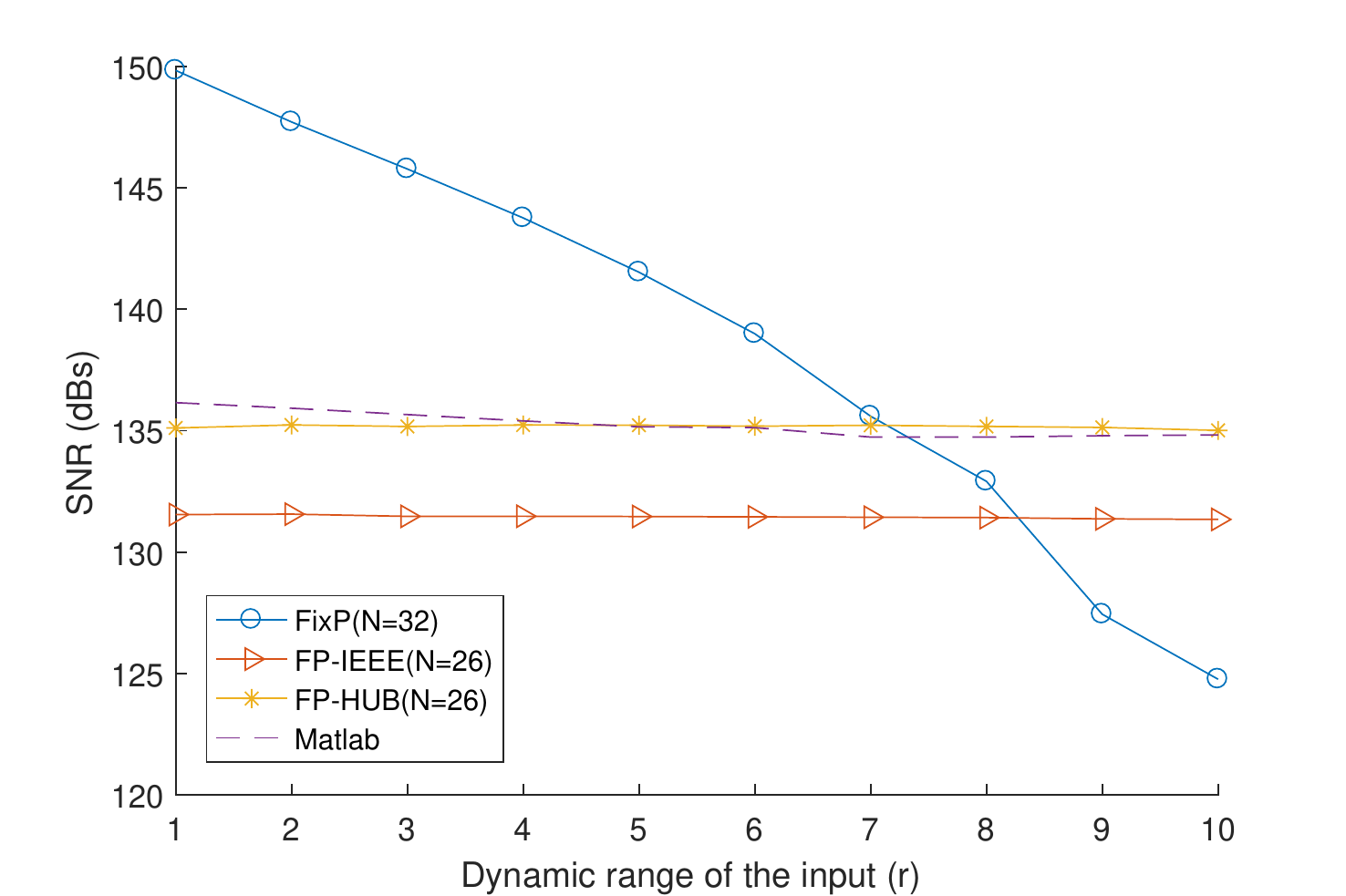}%
\label{fig-fixvsfp10}}
\caption{Precision of fixed- and floating- point approaches when varying $r$ (dynamic range of the input). }
\label{fig-fixvsfp}
\end{figure}

The main reason to use a FP implementation instead of a fixed-point one is the fact that the FP approach increases the dynamic range of inputs values supported keeping a reasonable accuracy.  To show that, we have performed a similar experiment as in Subsection~\ref{sec:error} for the fixed-point architecture described in~\cite{TCAS15} and the FP ones proposed in this paper varying  the dynamic-range parameter $r$  from 1 to 40 (i.e. the  magnitude of the values in the input matrices range from $2^{-r}$ to $2^r$). For each value of $r$, 10,000 input 4x4 matrices are generated randomly using double-precision FP. These input matrices are scaled and/or rounded to fit the corresponding input format in each tested architecture.  Fig.~\ref{fig-fixvsfp} shows the mean SNR of the results for both, fixed and floating point approaches. The fixed-point approach (FixP) has 32 bit-width and the FP ones use single-precision (32 bits for inputs and outputs)  and $N=26$ bits (the internal precision of the significand) for both IEEE and HUB approaches. The results using MatLab for single-precision is included as a reference. Fig.~\ref{fig-fixvsfp40} presents the results for all experiments whereas Fig.~\ref{fig-fixvsfp10} shows a zoom-in for $r$ up to 10. It is observed that the fixed-point approach produces better results than the FP ones for low values of the dynamic-range parameter $r$ (see Fig.~\ref{fig-fixvsfp10}). This is because the number of effective bits used for computation in FixP is larger than in the FP ones. However, this advantage decreases rapidly when $r$ increases, being the SNR of the FixP implementation lower than the FP-HUB one from $r=8$. As it is shown in Fig.~\ref{fig-fixvsfp40}, the SNR for FP approaches remains in reasonable values when $r$ increases while the SNR of FixP decreases steadily until $r$ reach 14 when it slumps. As a consequence, although a FP implementation is typically more expensive than a fixed-point one, for some specific applications, the use of FP may be compulsory due to the dynamic range of the input values.

To compare the implementation cost, we use the same fixed-point rotator that we used in the error analysis shown in Fig.~\ref{fig-fixvsfp} and the best of the FP ones (the HUB version). Note that the 32 bit-width fixed-point rotator utilizes 27 CORDIC iterations since that number of iterations offers the maximum precision for that bit-width; and the FP-HUB rotator with $N=26$ utilizes 24 CORDIC iterations. The fixed-point implementation results do not include the circuit for scaling the input and output values that this implementation may require.  Table~\ref{tab:fixvsFP} summarized the obtained implementation results. As expected, FP implementation requires more area than the fixed-point one, but this increase is slight, and even the number of registers utilized decreases. In contrast, the critical-path delay is lower for the FP rotator and, consequently, the throughput is increased in the same amount (since the number of cycles per rotation is the same for both approaches). However, we should point out that the latency of the FP rotator is slightly higher due to the input and output converters. Similarly, power consumption is significantly increased by the FP implementation, but that is mainly due to the increase in speed. If the frequency were set to the maximum frequency of the fixed-point approach, the power consumption of the FP approach would be 0.138 Watts, which is a very little increase compared to the fixed-point approach. On the other hand, even considering the 18\% of speed increase, the consumption increase in terms of energy is scarce.         

\begin{table}[th]
\caption{Fixed-point vs FP implementation results in Virtex-6}
    \label{tab:fixvsFP}   
\centering
    \begin{tabular}{llllll}
   \hline\noalign{\smallskip}
 Format & Delay& LUTs&Registers&Power&Energy\\
 \noalign{\smallskip}\hline\noalign{\smallskip}
FixP(32) & 3.26 ns & 1947  & 1914 & 0.132 W &430 pJ  \\ 
FPHUB 32(26)& 2.66 ns& 2182 & 1785&0.168 W& 448 pJ\\
\noalign{\smallskip}\hline\noalign{\smallskip}
FP/FixP (\%) &-18.4&	12.1&	-6.7&	27.3&	4.2 \\
\noalign{\smallskip}\hline

    \end{tabular}
   
\end{table}

Summarizing, if the target application has a low dynamic range, the fixed-point approach may provide the best precision and less hardware cost. However, for applications with higher dynamic range, FP architectures will provide much better precision with a slight increase of area and energy but with higher throughput.

\subsection{Comparison with previous FP implementations}

As we said before, to the best of our knowledge there is no previous hardware implementation of a specialized FP Givens rotation unit based on CORDIC. Therefore, we provide a comparison with circuits as similar to our proposal as we have found. Taking this into account, we should note that we can only provide a rough comparison and evaluation. Also notice that to provide comparable results, our designs have been re-synthesized using the same FPGA family as the ones in \cite{Munoz2010},\cite{4637696}, and \cite{Wang20093}, specifically Virtex-5 (XC5VLX330T-2). For the other designs, we use the data provided by the authors on those papers.   

The CORDIC co-processors in~\cite{4637696} and~\cite{Munoz2010} allow performing the Givens rotation carried out by our rotator, although they are not optimized for this purpose. Table~\ref{tab:Pcomparison} and Table~\ref{tab:Acomparison} summarize the comparison between these FP double-precision CORDICs and our HUB rotator using the same precision and technology. The circuit to control the rotation operation or to store temporal values for the CORDIC processor approaches are not considered in these results. The initiation interval (i.e., the minimum number of cycles between two consecutive rotations) is represented depending on $e$, which is the number of elements in each row. Conversely, the throughput at maximum supported frequency is calculated considering an example with 8 elements per row (the same size used in the error analysis) to facilitate comparison. This throughput is expressed in millions of Givens rotations per second. 

% \begin{table*}[thb]
% \renewcommand{\arraystretch}{1.2}
% \caption{Comparison between similar designs on Virtex-5}
% \label{tab:comparison}
% \centering
% \begin{tabular}{|c|cccccccccc|}
% \hline
% Design& Precision&Max Freq & Latency& Initiation Interval & Throughput  & LUTs&Registers&Slices&DSPs&BRAM\\
%       &          & (MHz)   & (cycles)&  (cycles)               &  (MOp/s)&     &          &      &    &   \\\hline
% FP CORDIC (\cite{Munoz2010})&Double& 67.1&224 &$212+e\times224$& 0.033 (e=8) & 11,718&600&- &0&0\\ %4 cyclos por iteracion +12 Falta las multiplicaciones 

% FP CORDIC (\cite{4637696})&Double&173.3&69x2&$69+e\times1$& 2.25 (e=8) & 22,189&20,443&-&0&0\\
% HUB FP rotator &Double &255.8& 60&$e\times1$ & 31.97 (e=8) &8,463&7,598&-&0&0\\\hline  %($N=58$ y 55 iteraciones)
% 7x7 FP QRD (\cite{Wang20093})&Single &132.0& 954& 364&0.36 &-&-&126,585&102&56\\
% Our 7x7 HUB FP QRD& Single&287.8 &296 &7 &41.11 &- &- &50,547&52 &0\\\hline
% \end{tabular}

% \end{table*}

\begin{table}[thb]
\caption{Performance comparison among similar designs on Virtex-5}
\label{tab:Pcomparison}
\centering
\begin{tabular}{lllll}
\hline\noalign{\smallskip}
Design&Max Freq & Latency& Initiation Interval & Throughput  \\
              & (MHz)   & (cycles)&  (cycles)               &  (MOp/s) \\
             \noalign{\smallskip}\hline\noalign{\smallskip}
FP CORDIC (\cite{Munoz2010})& 67.1&224 &$212+e\times224$& 0.033 (e=8) \\ %4 cyclos por iteracion +12 Falta las multiplicaciones 
FP CORDIC (\cite{4637696})&173.3&69x2&$69+e\times1$& 2.25 (e=8) \\
HUB FP rotator  &255.8& 60&$e\times1$ & 31.97 (e=8) \\%($N=58$ y 55 iteraciones)
\noalign{\smallskip}\hline\noalign{\smallskip}  
7x7 FP QRD (\cite{Wang20093}) &132.0& 954& 364&0.36\\
Our 7x7 HUB FP QRD& 287.8 &296 &7 &41.11 \\
\noalign{\smallskip}\hline
\end{tabular}
\end{table}

\begin{table}[thb]
\caption{Area comparison among similar designs on Virtex-5}
\label{tab:Acomparison}
\centering
\begin{tabular}{lllllll}
\hline\noalign{\smallskip}
Design& Precision & LUTs&Registers&Slices&DSPs&BRAM\\
      \noalign{\smallskip}\hline\noalign{\smallskip}
FP CORDIC (\cite{Munoz2010})&Double& 11,718&600&- &0&0\\ %4 cyclos por iteracion +12 Falta las multiplicaciones 

FP CORDIC (\cite{4637696})&Double& 22,189&20,443&-&0&0\\
HUB FP rotator&Double  &8,463&7,598&-&0&0\\%($N=58$ y 55 iteraciones)
\noalign{\smallskip}\hline\noalign{\smallskip}
7x7 FP QRD (\cite{Wang20093})& Single &-&-&126,585&102&56\\
Our 7x7 HUB FP QRD& Single &- &- &50,547&52 &0\\
\noalign{\smallskip}\hline
\end{tabular}

\end{table}

Being a fully pipelined design, as expected, the main advantage of our rotator is its throughput. Considering 4x4 matrices ($e=8$), the throughput of our HUB rotator is 15 times higher than for~\cite{4637696} and three orders of magnitude higher than for~\cite{Munoz2010} (see Table~\ref{tab:Pcomparison}). This is because both generic CORDICs need to perform completely the angle calculation before starting to rotate the row elements. Hence, \cite{4637696} could not take full advantage of its pipeline implementation and it could produce at most one Givens rotation each $(69+e)$ cycles. In contrast, our design could perform a rotation each $e$ cycles. Consequently, even considering the same frequency, the difference is very significant for small matrices. For larger matrices, the relative difference would be reduced but our design always will have higher throughput than~\cite{4637696}.  Furthermore, our rotator requires almost a third of the FPGA resources used by~\cite{4637696} (see Table~\ref{tab:Acomparison}) and the latency is less than half.

If we consider~\cite{Munoz2010}, the difference is beyond comparison. Since \cite{Munoz2010} is a word-serial implementation, it uses less than 10 times less register than our pipeline architecture, but surprisingly a 40\% more LUTS. Similarly, this word-serial nature produces that the throughput of \cite{Munoz2010} is several orders of magnitude smaller than ours for small matrices, and this difference would increase dramatically if the size of the matrices increases. However, we must remember that the design in~\cite{4637696} can compute many other elementary functions, and our design only computes Givens rotations. 

On the other side, Table~\ref{tab:Pcomparison} and Table~\ref{tab:Acomparison} show also a comparison with the QRD calculator presented in~\cite{Wang20093} (see Section~\ref{sec:review}). In~\cite{Wang20093}, authors provides implementation results for a FP single-precision QRD calculator for 7x7 matrices. We have calculated the cost of implementing an equivalent QRD calculator using the HUB version of our Givens rotator for the same technology (Virtex-5)  configured with the architecture proposed in~\cite{TCAS15}. Table~\ref{tab:Acomparison} shows that our design utilizes less than half of the resources utilized by the architecture proposed in~\cite{Wang20093}, even without counting the BRAMs. More importantly, Table~\ref{tab:Pcomparison} shows that using the maximum frequency supported for each circuit, our design has six times less latency and computes 100 times more matrices per second.  This fact reinforces the idea that the CORDIC approach is the best way of implementing Givens rotations in hardware. 

\section{Conclusion}\label{sec:con}
In this paper, we propose a very effective hardware design of a Givens rotation unit for floating-point computation based on the CORDIC algorithm. As in previous FP CORDIC architectures proposals, the FP Givens rotator is based on a fixed-point one with input and output converters.    We provide a detailed description of two different approaches, one for conventional FP formats and another for new HUB formats. The error analysis and the FPGA implementation results reveal that the proposed FP units expand the dynamic range of inputs values and require only a moderate increase of hardware utilization compared to previous fixed-point one. Furthermore, the HUB approach significantly improves the area, delay, and energy consumption of the conventional one. Comparison with other FP units to compute QRD shows that using the proposed design to compute QRD improves the throughput, more than one order of magnitude, and simultaneously reduces the area by more than half. The proposed units could be used to design both highly parallel QRD units and low-cost iterative ones.

\begin{acknowledgements}
This work was supported in part by following Spanish projects: TIN2016-80920-R, and JA2012 P12-TIC-1692.
\end{acknowledgements}


 \end{document}